\documentstyle[preprint,aps,psfig]{revtex}

\newcommand{\be}{\begin{eqnarray}}
\newcommand{\ee}{\end{eqnarray}}
\newcommand{\bc}{\begin{center}}
\newcommand{\ec}{\end{center}}

\begin{document}
\draft
\tighten

\title {THE RELATIVISTIC ELASTICITY OF RIGID BODIES}
\author{ A. Brotas\footnote{E-mail:brotas@fisica.ist.utl.pt} ,
    J.C. Fernandes\footnote{E-mail:joao.carlos@tagus.ist.utl.pt},   }
\address{ Departamento de F\'{\i}sica, Instituto Superior T\'ecnico, \\
Av Rovisco Pais 1096.  Lisboa Codex, Portugal}

\date{\today}

\maketitle

\begin{abstract}

	In 1909 Born studied the {\it relativistic undeformable body} but 
made the mistake of calling it {\it rigid}. The "{\it rigid body}" as one can find 
in Relativity books is, in fact, this Born {\it undeformable body}.
	In Relativity it is necessary to distinguish between {\it rigid} and 
{\it undeformable}.  The {\it rigid} body (in the sense of the most rigid possible) 
must be the {\it deformable} body where schock waves propagate with maximum speed $c$. 

	The elastic laws of rigid bodies are :

\bc $p = \frac{\rho_0^0 c^2}{2} \left( \frac{1}{s^2} - 1 \right) ~~~~;~~~~ 
 \rho_0 = \frac{\rho_0^0}{2} \left( \frac{1}{s^2} + 1 \right) $ \ec
\bc with: $ s = \frac{l}{l_0} = 1 - \frac{\Delta l}{l_0} = 1 - \delta $ \ec
\bc $\rho_0^0$ is the body density (not compressed) in proper system. \ec
  
 We think that these laws, which are ignored by the majority of relativists,
should be taught in the elementary relativistic courses.

	With the approach of 2005, the centenary year of Relativity, we should
like to appeal to all those who have some influence on these matters to 
avoid this mistake of repeatedly calling {\it rigid} to the {\it undeformable body}.

\end{abstract}
\newpage

%--------------------------------------------

\section{An old problem}

	In 1909 Born studied the relativistic {\it undeformable body} but made the mistake 
of calling it {\it rigid} \cite{um}. The "{\it rigid body}" we can find in a number of
Relativity texts \cite{dois} is, in fact, this Born {\it undeformable body}. 
 This mistake gave rise to a number of paradoxes \footnote[1] 
{Paradox of the rotating disk; paradox of the three degrees of freedom of "rigid" 
bodies in Relativity; paradox of the angle lever; difficulties with problem of heat 
transmission in Relativity; paradox of the black hole fishing, etc..} 
 that, even today, many relativists are unable to solve 
 but that can be easily clarified noting that 
in Relativity the two concepts {\it rigid} and {\it undeformable} must be separated. 
The {\it rigid body}, in the sense of {\it as rigid as possible}, must be in Relativity 
the {\it deformable} body where shock waves propagate with maximum speed $c$.

	Elastic laws for relativistic rigid bodies (one dimension) where discovered 
by McCrea \cite{crea}, in 1952, and later rediscovered and generalized, to 2 and 3 
dimensions in 1968, 
by one of the authors of this text \cite{brotas1}. These laws, ignored untill today by 
the great majority of relativists, are very curious and may (we think they 
should) be taught in elementary courses. They allow us to answer some students' 
questions (usually without answer) about what could happen under some circumstances.

{\bf	Now that year 2005 is approaching and will be celebrated the first 
centenary of Relativity, we appeal all those who can have influence 
in this matter so that in Relativity books to be published by this centenary, 
the mistake of calling {\it rigid} to the {\it undeformable body} do not persist
and the relativistic elastic laws do not be ignored. }

	We present in this text two deductions of these laws for one dimension
that can only surprise because they are, at the same time, unknown and very easy.

%--------------------------------------------

\section{Preliminary note}

	 Bodies, in reality, are more or less rigid. In Physics books, however, 
 the word {\it rigid} is allways understood as "{\it the most rigid possible}".
In Classical Physics nothing prevents us from conceiving these 
{\it limit-rigid} bodies as undeformable, i.e. bodies keeping the 
same form independently of their movements or forces acting on them. 

	Although the notion of {\it rigid} is a physics concept and the notion 
of {\it undeformable} 
is a geometrical one, they are thus accepted as synonymous in Classical 
Physics books. The elasticity modulus of these {\it rigid-undeformable 
bodies} is infinite and the shock waves propagate on them with infinite 
velocity. 

	In Relativity, we do not accept  material 
waves that can transmit energy or signals with a velocity greater than 
$c$. So we can't accept the existence of bodies where these 
waves are possible. The {\it relativistic rigid body} must be 
 the body where waves (and in particular shock waves) 
propagate with maximum speed $c$.

	We will show that this simple property leeds us to
the elastic laws for these bodies.

%--------------------------------------------

\section{The velocity of shock waves and the elastic laws}
%--------------------------------------------

\subsection{In Classical Physics}

	Let us consider an homogenous bar of section $S$, length 
$l_{0}$ and density $\rho_{0}$ (when not deformed), moving longitudinally 
with velocity $v$ that at the instant $t=0$ strikes an undeformable 
wall (or an equal bar coming from the opposite side with velocity $-v$). \\

	Let us suppose that, after the impact, a shock wave with a front 
 velocity $V_p$ (in the wall frame), will propagate along the bar, 
 separating the fraction of the bar stopped and uniformly compressed, from 
another still in movement and not compressed. 

	The wave front reaches the end edge at instant:
\be  t_1 = \frac{l_{0}}{v + V_p} \ee

At that moment $t_1$ the bar is wholy stopped and compressed, and has length
\footnote[2] 
{In the sequence, in the interval $\left[ t_1 , 2t_1\right]$, a new 
wave will propagate in the bar with velocity $V_p$, but in the opposite 
way, separating a new fraction not compressed that turns back with 
velocity $-v$, from the compressed fraction still fixed. At the instant 
$t=2t_1$, the all bar is again in movement with velocity $-v$. }:

\be l_1 = t_1 V_p =  \frac{l_{0} V_p}{v + V_p} \ee

   The wave front velocity in the moving bar system is: 
\bc $V_b = v + V_p $ \ec

 \bc We will use notation: \ec
\bc \[ \Delta = l_0 - l_1 = \frac{l_0 v}{v + V_p} = \frac{l_0 v}{V_b} \] \ec
\bc \[ \delta = \frac{\Delta}{l_0} = 
\frac{v}{v + V_p} = \frac{v}{V_b}, ~~~~~ s = \frac{l_1}{l_0} = 
\frac{l_0 - \Delta}{l_0} = 1 - \delta    \] \ec

	The kinetic energy of the moving bar, given by,
\be E_c = \frac{1}{2} l_0 S \rho_0 v^2 \ee
must be equal to the elastic energy of the bar stopped and compressed, 
given by:
\be E_{el} = - \int_{l_0}^{l_1} S p dl  = \int_{0}^{\delta} S p l_0 d\delta \ee 

	This equality must be verified no matter the velocity $v$ may be. 
 Let us suppose that the front wave velocity $V_b$ is independent of the value of $v$.
	Equating the two energies and differentiating in order to $v$ 
one obtains:
\be l_0 S \rho_0 v = S p l_0 \frac{1}{V_b} \ee 
This allows us to write:
\be p = \rho_0 v V_b = \delta \rho_0 V_b^2 \ee

  This calculation indicates that 
 the bar material elasticity, is linear. In other words, its a Hook material
 \footnote[3] 
{In the study of a bar compression, we have: 
$\Delta > 0$, $\delta > 0$, $s < 1$ and find a positive $p$ corresponding 
to a compression for every positive $\delta$.
	To study traction, one can imagine that the opposite edge of the bar 
is stopped at instant $t=0$ by some device. In this case, the bar front 
goes on with velocity $v$. The bar stops completely only at instant $t_1$ 
having at that moment the length $l_1$, being:
$ t_1 = \frac{l_0}{-v+V_p} ~~~~ ; ~~~~  l_1 = \frac{l_0V_p}{-v+V_p} $
All preceding formulae remain valid if we use $-v$ instead of $v$. In this 
case: $\Delta < 0$, $\delta < 0$, $s > 1$ and find a negative $p$, 
corresponding to a traction for every negative $\delta $.}. 
	But a confirmation is required. We must verify that this result 
agrees with the one obtained from the study of linear momentum change.
. \\

	During the time period $\left[ 0 , t_1\right]$ the change of bar 
momentum is:
\be \Delta P = 0 - \rho_0 S l_0 v \ee

	For this period, the wall acts 
on the bar with a force $F = - p S$ and transmits it the impulse:
\be I = - S ~ p ~ t_1 ~~ = ~~ - S p \frac{l_0}{V_b} \ee

Equating $I$ with $\Delta P$ we find, in this way, the same result found 
in previous calculation (equation 6).
	The hypothesis that the front wave velocity $V_b$ is independent 
of $v$ is then acceptable, since it leads to a result 
compatible with the two major conservation principles of Physics 
\footnote[4] 
{ If the hypothesis used were to keep $V_p$ constant (for every $v$) 
instead of $V_b$, the calculations made from energy conservation and 
change of momentum would lead to different elastic laws, which means 
the bar had not the behaviour we supposed.}.

    With the usual notation $p = \delta ~ E = (1-s)E$, the elasticity 
modulus $E$ of the bar material is given by:
\be E = \rho_0 V_b^2 \ee
and allows us to  write:
\be V_b = \sqrt{\frac{E}{\rho_0}} \ee
a result in agreement with the one predicted by the Alembert equation:
\be \frac{\partial^2x}{\partial X^2} - \frac{\rho_0}{E} 
\frac{\partial^2 x}{\partial t^2} = 0 \ee
%--------------------------------------------

\subsection{In Relativity}
	
	To obtain the relativistic elastic laws of a material from the 
velocity of schock waves, we use a procedure similar to the previous, 
taking into account naturally, the differences imposed by the theory.

\begin{figure}
\centerline{
	\parbox[t]{4.0in}{
	\psfig{file=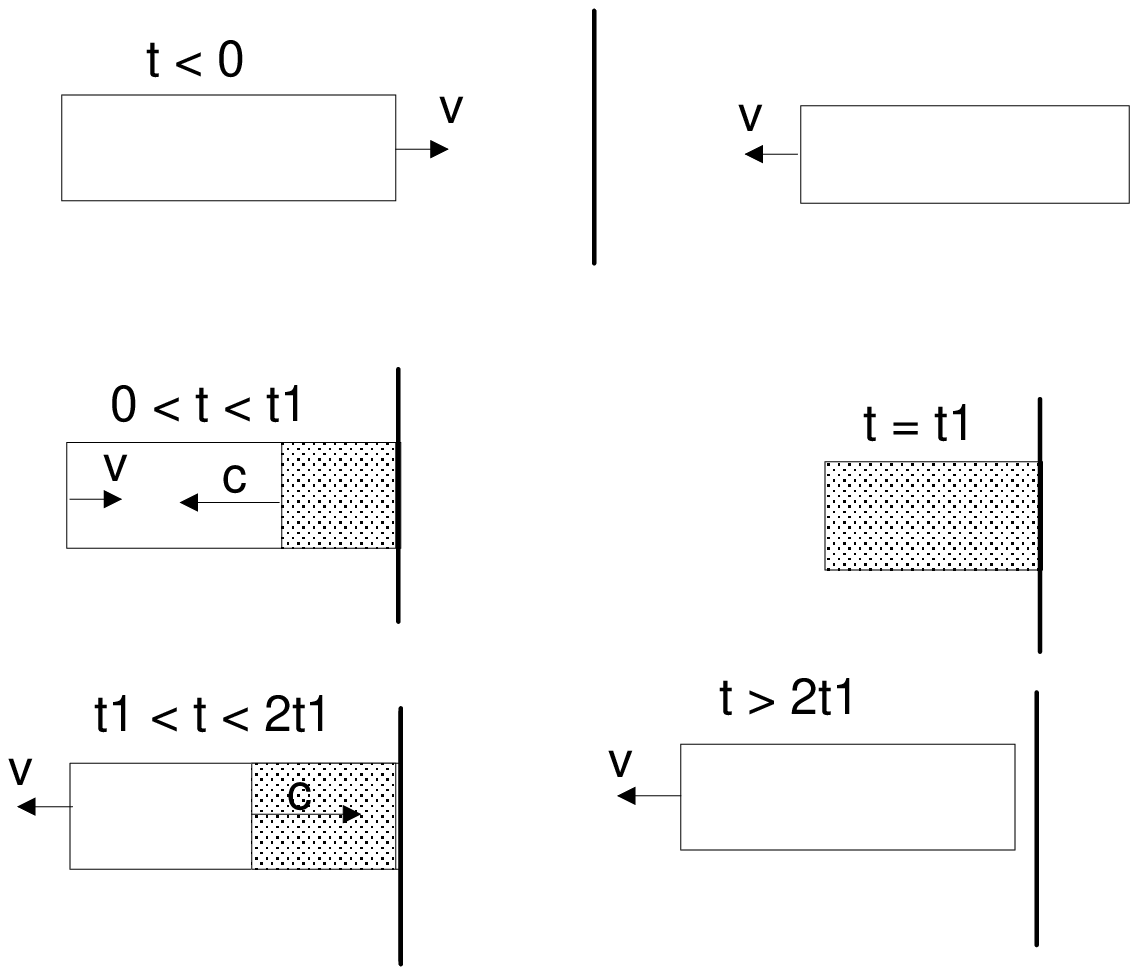,height=3.8in,width=3.8in}}
}
	\caption{Shock of a bar against an undeformable wall, or an equal bar
coming from the opposite direction.}
	\label{fig:fig2}
\end{figure}

	Let us consider a bar where a shock wave has a velocity $V = c$ 
(fig. \ref{fig:fig2}).
 	In this case, due to the relativistic law for 
velocity composition, we have:

\bc $ V_b = V_p = c $ \ec

In the wall frame, the bar is moving with velocity $v$ and it's length is:

\be l_v = l_o \sqrt{1-\beta^2} , ~~~~ with ~~~~\beta = \frac{v}{c} \ee 

	The bar is stopped and wholy compressed at instant,
 
\be t_1 = \frac{l_v}{v+c} = \frac{l_0}{v+c} \sqrt{1-\beta^2} \ee 
and its length is in that moment:

\be l_1 = l_0\sqrt{1-\beta^2}\frac{c}{v+c} = 
    l_0 \sqrt{\frac{1-\beta}{1+\beta}} \ee 

Using previous notations, we have in this case : 
\be s = 1-\delta = \frac{l_1}{l_0} =  \sqrt{\frac{1-\beta}{1+\beta}} \ee 

\bc We will use the following relations: \\ 

\[ 
 \frac{1}{s} + s ~~~ = ~~~\frac{2}{\sqrt{1-\beta^2}} ; ~~~~~ 
      \beta ~~ = ~~ \frac{1-s^2}{1+s^2} ; ~~~~~ t_1 ~~ = ~~ \frac{l_0}{c}s  
 \] \ec
 
The bar kinetic energy in movement is: 
\be E_c = S \rho^0_0 l_o c^2 \left[ \frac{1}{\sqrt{1-\beta^2}} -1 \right] = 
   \frac{S \rho^0_0 l_o c^2}{2} \left[ \frac{1}{s} + s - 2 \right] \ee 
Where $\rho^0_0 $ is the bar density (not compressed) in its proper system. 

The elastic energy of the compressed bar is given by a formula equal to 
the one in Classical Physics:

\be  E_{el} = - \int_{l_0}^{l_1} S p dl  = \int_{0}^{\delta} S p l_0 d\delta 
   = - \int^{s}_{1} S p l_0 ds \ee 

These two energies must have the same value. These equality must be 
verified no matter the velocity $v$ and $s$ may be. 

	Equating the two energies and differentiating with respect to $s$ one obtains:

\be \frac{S \rho^0_0 l_o c^2}{2} \left[ \frac{-1}{s^2} + 1 \right] 
    = - S p l_o \ee 
and: 
\be p = \frac{\rho^0_0 c^2}{2} \left( \frac{1}{s^2} - 1 \right) \ee 

	On the other hand, the change of momentum during the period $[0,t_1]$:
 
\be \Delta P = 0 - \frac{S \rho^0_0 l_o v}{\sqrt{1-\beta^2}} = 
  - \frac{S \rho^0_0 l_o c \beta}{2} \left[ \frac{1}{s} + s \right] \ee 
must be equal to the impulse:
\be I = - S ~ p ~ t_1 = \frac{S p l_0}{c} s \ee
allowing us to write:

\be p = \frac{\rho^0_0 c^2}{2} \left( \frac{1-s^2}{1+s^2} \right) 
    \left(\frac{1+s^2}{s^2} \right) = \frac{\rho^0_0 c^2}{2} \left( 
    \frac{1}{s^2} - 1 \right) \ee

\begin{figure}[h!]
\centerline{
	\parbox[t]{4.0in}{
	\psfig{file=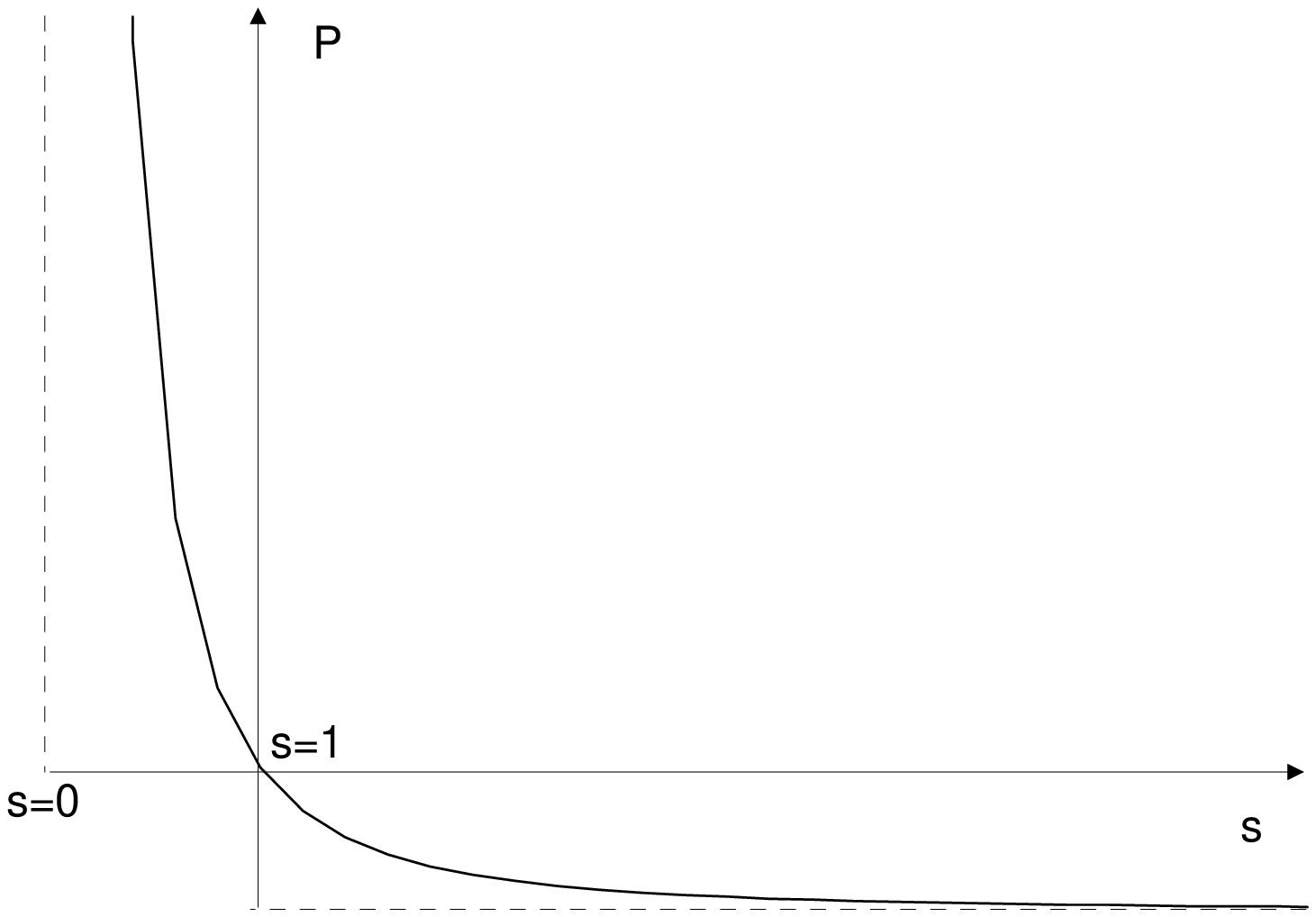,height=3.0in,width=3.0in}}
}
	\caption{Relativistic elastic law. \\
 We plot the pressure as function of deformation: \\  
 $p > 0$, compression, maximum value of $p = \infty$ for $s = 0$ ; \\
 $p < 0$, traction, maximum value $- \frac{\rho^0_0 c^2}{2}$ for $s = \infty$. }
	\label{fig:figura3}
\end{figure}

	We thus find the same elastic law (fig. \ref{fig:figura3}), using both principles. \\

	In the relativistic case the material density as a function of 
the deformation must be studied. We represent by  $\rho_0$ the density 
of the deformed material in its proper system.

	This problem is not, as it was in Classical Physics, a 
problem of accumulation of the same mass in a different volume since, 
when we compress the bar, we are supplying energy and so increasing 
its mass. Taking into account this increase, the density $\rho_0$ is: 

\be \rho_0 = \frac{E_c}{c^2 S l_1}  = \frac{m_0}{S l_1 \sqrt{1 - \beta^2}}
 = \frac{l_0 \rho_0^0}{l_1 \sqrt{1 - \beta^2}}   \ee 
allowing to write: \footnote[4] 
{ It does not exist to our knowledge any book where these formulae (19) and (24) 
are written. Nevertheless, eliminating $s$ between them, one obtains the formula:
\bc $\rho_0 = \rho^0_0 + \frac{p}{c^2} $ \ec 
that can be found in several treatises in chapters about relativistic fluids. 
This result holds in the case of fluids where waves propagate with speed $c$. 
If fluids are confined in cylinders, the variations of pressure 
just imply change on its length. So the fluids to which this formula 
applies are "rigid" fluids, in the sense that they are deformable but 
they have maximum incompressibility consistent with Relativity }: 
\be \rho_0 = \frac{\rho^0_0}{2} \left( \frac{1}{s^2} + 1 \right) \ee 

	It is easy, having in mind the momentum variation of the 
different elements, to write the equation of motion for a bar whose 
material obeys these laws:

\be  \frac{\partial^2X}{\partial x^2} - \frac{1}{c^2} 
   \frac{\partial^2 X}{\partial t^2} = 0 \ee

	We notice that this equation is invariant under a Lorentz 
transformation and its solutions are written in Euler representation:
 $X=X(x,t)$, while the solutions of the traditional Alembert equation, 
 invariant under a Galileo transformation, are written in Lagrange 
representation : $x=x(X,t)$. For this, equation (25) must not be looked 
as corresponding to Alembert's equation (11) when $V = c$, neither as 
its limit case when $V \rightarrow \infty $. To this limit case correspond 
the undeformable bodies that have no existence in Relativity

%--------------------------------------------

\section{Last note}

	In 1985 Luis Bento published the three-dimensional equation of 
motion for a rigid material and shaw that the velocity of transversal 
waves is $v_t = \frac{c}{\sqrt{2}}$, in the case of a null Poisson 
coefficient.

	Generalization to one dimension for a hook material "non rigid", 
was published in 1980. \cite{brotas2}
	However, did not come to our knowledge, the generalization to 
2 and 3 dimensions for the non-isotropic materials case. In other words, 
we don't know macroscopic relativistic models to describe the 
mechanical behaviour of crystals, but will be certainly possible to find 
out these models because crystals exist! (even if we don't find them, 
there is always the possibility of predicting theoretically their existence).

%--------------------------------------------

%--------------------------------------------

\end{document}